\documentclass[preprint, times, 5p]{elsarticle}
\usepackage{subcaption}
\usepackage{mhchem}
\usepackage{lineno}
\usepackage{amsmath}
\usepackage{siunitx}
\usepackage{threeparttable}
\usepackage{derivative}
\usepackage{natbib}
\usepackage{graphicx}

\modulolinenumbers[1]
\usepackage{docmute}
\usepackage[colorlinks=true]{hyperref}

\newcommand*\patchAmsMathEnvironmentForLineno[1]{
  \expandafter\let\csname old#1\expandafter\endcsname\csname #1\endcsname
  \expandafter\let\csname oldend#1\expandafter\endcsname\csname end#1\endcsname
  \renewenvironment{#1}
  {\linenomath\csname old#1\endcsname}
  {\csname oldend#1\endcsname\endlinenomath}}
\newcommand*\patchBothAmsMathEnvironmentsForLineno[1]{
  \patchAmsMathEnvironmentForLineno{#1}
  \patchAmsMathEnvironmentForLineno{#1*}}
\AtBeginDocument{
  \patchBothAmsMathEnvironmentsForLineno{equation}
  \patchBothAmsMathEnvironmentsForLineno{align}
  \patchBothAmsMathEnvironmentsForLineno{flalign}
  \patchBothAmsMathEnvironmentsForLineno{alignat}
  \patchBothAmsMathEnvironmentsForLineno{gather}
  \patchBothAmsMathEnvironmentsForLineno{multline}
}

\newcommand{\dedx}{\odv{E}{x}}
\newcommand{\dedxinline}{\odv[switch-/=true]{E}{x}}
\newcommand{\comptonsoft}{ComptonSoft}
\newcommand{\nq}{N_\mathrm{q}}
\newcommand{\nph}{N_\mathrm{ph}}
\newcommand{\nel}{N_\mathrm{e}}

\begin{document}
\begin{frontmatter}
	\title{Charge Response Simulation of Liquid Argon Compton Cameras for Precise MeV Gamma-Ray Event Reconstruction}
	\author[utokyo,uosaka]{Shota Arai\corref{cor1}}
	\author[uosaka,ipmu]{Hirokazu Odaka}
	\author[riken,uosaka]{Satoshi Takashima}
	\author[uosaka]{Kentaro Shirahama}
	\author[uosaka,utokyo]{Masashi Tanaka}
	\author[uosaka]{Ryutaro Tatsumi}
	\author[uosaka]{Hodaka Kawamura}
	\author[utokyo]{Kouichi Hagino}
	\author[utokyo,resceu,tsqsi]{Aya Bamba}

	\cortext[cor1]{Email Address: shota.arai@phys.s.u-tokyo.ac.jp}

	\address[utokyo]{Department of Physics, Faculty of Science, The University of Tokyo, 7-3-1 Hongo, Bunkyo-ku, Tokyo 113-0033, Japan}
	\address[uosaka]{Department of Earth and Space Science, The University of Osaka, 1-1 Machikaneyama-cho, Toyonaka, Osaka 560-0043, Japan}
	\address[ipmu]{Kavli IPMU, The University of Tokyo, 5-1-5 Kashiwanoha, Kashiwa, Chiba 277-8583, Japan}
	\address[riken]{RIKEN Nishina Center, 2-1 Hirosawa, Wako, Saitama 351-0198, Japan}
	\address[resceu]{Research Center for Early Universe, Faculty of Science, The University of Tokyo, 7-3-1 Hongo, Bunkyo-ku, Tokyo 113-0033, Japan}
	\address[tsqsi]{Trans-Scale Quantum Science Institute, The University of Tokyo, 7-3-1 Hongo, Bunkyo-ku, Tokyo 113-0033, Japan}

	\begin{abstract}
		We studied the detector response for a liquid-argon time-projection chamber (LArTPC) Compton camera targeting MeV gamma-ray astronomy by a Geant4-based Monte Carlo simulation. The simulation, implemented within ComptonSoft, incorporates electron-ion recombination, charge diffusion, and induction on the electrode pixels. We evaluated the angular resolution measure (ARM) and the  energy reconstruction error as functions of electronic noise, pixel pitch, and the recombination correction method. An electronic noise level of \qty{4}{\keV} and a pixel pitch of \qty{1}{mm} yield an ARM (FWHM) of about \qty{3}{\degree} at \qty{1332.5}{\kilo\electronvolt} and an energy reconstruction error better than \qty{4}{\percent} for incident energies of \qty{1173.2}{\keV} and above. Recombination correction using both scintillation light and charge signals improves the ARM and the energy error by factors of 1.7--2.3 and 2.2--5.4, respectively, compared to the charge-only correction, although the uncertainty of the recombination-fluctuation model remains. Since this simulation relies on idealized assumptions such as the use of the Monte Carlo truth in the event reconstruction, these values represent the best cases achievable under these detector conditions; nevertheless, they show that a large-scale LArTPC is a promising Compton camera for cosmic MeV gamma-ray observations.
	\end{abstract}

\end{frontmatter}
\section{Introduction}
	MeV gamma rays serve as a unique probe of high-energy astrophysical processes. Nuclear de-excitation and radioactive decay lines in the MeV band provide direct signatures of nucleosynthesis and low-energy cosmic rays. In addition, gamma rays in a broader energy range including the MeV band are produced through pion decay, synchrotron radiation, and inverse Compton scattering by high-energy hadrons and leptons, making them a key diagnostic for identifying cosmic-ray acceleration sites such as supernova remnants and the Galactic Center region.

	Despite this scientific importance, the MeV band is still poorly explored because existing detectors have limited effective areas. This situation is often referred to as the ``MeV gap'' \citep{GRAMS2020}.
	The COMPTEL detector onboard the Compton Gamma-Ray Observatory, launched in 1990, achieved remarkable results, including the detection of the \qty{1.809}{MeV} line from \ce{^26Al}, yet catalogued only 63 sources during its 10-year observation \citep{COMPTEL}.

	To address this, a liquid-argon time-projection chamber (LArTPC) is considered a next-generation detector technology for MeV gamma-ray observations because it is cost-effective, scalable, and can be built with almost no dead volume, enabling a large effective area. In addition, its capability as a 3D particle tracker offers strong discrimination between gamma rays and the background. Several LArTPC-based Compton camera concepts, such as GRAMS \citep{GRAMS2020} and GammaTPC \citep{GammaTPC2025}, have been proposed.

	Although LArTPC technology is widely used in particle physics, it has not yet been broadly adopted for MeV gamma-ray astronomy, mainly due to technical challenges.
	Since Compton cameras use both the position and energy information to reconstruct events, good energy and position resolutions are required of the detector.
	In realistic conditions, several effects in liquid argon degrade these resolutions. Estimating their impact on the energy reconstruction performance is non-trivial.
	An actual detector is also subject to further uncertainties, such as calibration uncertainties and misidentification in the event reconstruction. These are, however, expected to be reduced as the technology and analysis methods mature, whereas the physical processes in liquid argon set a fundamental limit on the achievable performance. Clarifying this limit through detector-response simulations is therefore essential to define the development goals of an LArTPC Compton camera. In this study, we performed a Geant4-based simulation incorporating the key processes in liquid argon.

	The structure of this paper is as follows: Section \ref{sec:detector} describes how an LArTPC functions as a Compton camera. In Section \ref{sec:method}, we describe our implementation and method to simulate the response of an LArTPC. The data processing after simulation is described in Section \ref{sec:data_process}. In Section \ref{sec:ResultsAndDiscussion}, we evaluate the detector performance, namely the ARM and the energy reconstruction error as functions of the electronic noise level and the pixel pitch. We give the conclusion of this work in Section \ref{sec:conclusion}.
\section{Detector Concept of Liquid Argon Time-Projection Chamber as a Compton Camera}\label{sec:detector}
	Gamma rays interact with electrons bound to argon atoms via Compton scattering and photoabsorption. These interactions simultaneously produce ionization electrons and scintillation light of \qty{128}{\nano\metre}.

\begin{figure}
		\centering
		\includegraphics[page=4, keepaspectratio, width=\linewidth]{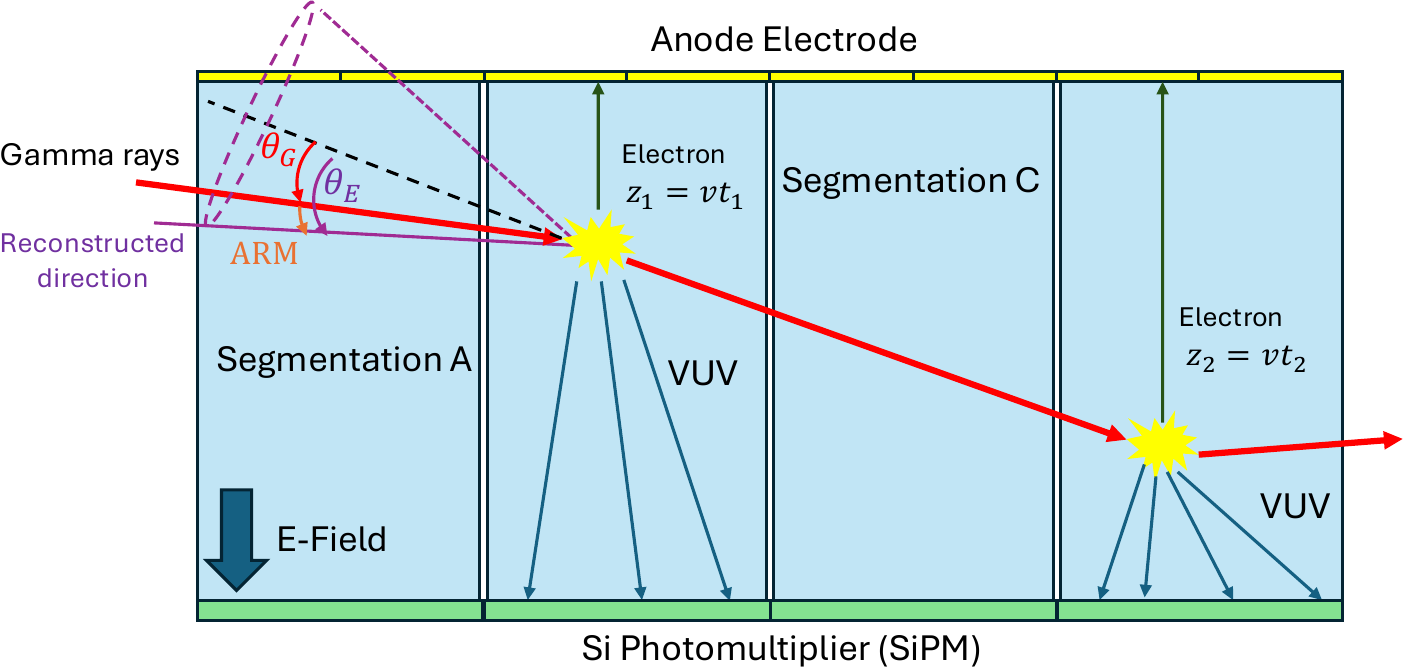}
		\caption{Schematics of the LArTPC. The entire active volume is filled with LAr. The silicon photomultiplier reads the scintillation light, and ionization electrons are drifted by the electric field and are collected at the anode electrode.} \label{fig:LArTPC}
	\end{figure}

	Figure~\ref{fig:LArTPC} shows a schematic of the detector. An LArTPC detects both charge and light signals, measuring the 3D coordinates of the interaction points and their deposited energies. The $z$ coordinate is determined from the difference between the detection times of the light and charge signals, while the $xy$ coordinates are derived from the detection point on the segmented anode. The energy deposition is obtained by the amount of the charge signal. Using this information for events with multiple hits, Compton event reconstruction can be applied to determine the direction and energy of the incident gamma-ray photon (see Section \ref{subsec:event_reconstruction}). If two or more events occur within one drift time, the association between the light and charge signals becomes ambiguous, and the $z$ coordinate cannot be determined correctly (pileup events), which is a serious problem in a high-background environment such as in space.
	In addition, since the scintillation light produced anywhere in the volume illuminates the whole detector, the light from individual interactions cannot be measured separately. To address these issues, a segmented configuration has been proposed \citep{GRAMS2020,GammaTPC2025}. The segmentation serves two purposes: it reduces the effective pileup rate,	because each segment can be treated independently; and it allows the scintillation light of each interaction to be measured separately, which is required for the recombination correction described in Section~\ref{subsec:energyCorrection}.

	To evaluate the performance of Compton cameras, the Angular Resolution Measure (ARM) is commonly used as a figure of merit. In this paper, we define the ARM as $\theta_E-\theta_G$, where $\theta_E$ is the Compton scattering angle derived by Compton kinematics, and $\theta_G$ is that derived from the geometrical information (Figure~\ref{fig:LArTPC}).

\section{Simulation Method}\label{sec:method}

	To simulate the detector response of an LArTPC, we used \comptonsoft\ \citep{ODAKA2010303}, a Compton camera simulation framework based on Geant4 \citep{ALLISON2016186}. It was used to construct the detector response of the Soft Gamma-ray Detector onboard the ASTRO-H observatory, and its flexible design allows detector-specific responses	to be implemented. \comptonsoft\  calculates several processes affecting the detector response; in this work, we implemented the processes described in Section \ref{sec:simulation:processes} into \comptonsoft.

\begin{figure*}
		\centering
		\includegraphics[page=3, keepaspectratio, width=0.9\linewidth]{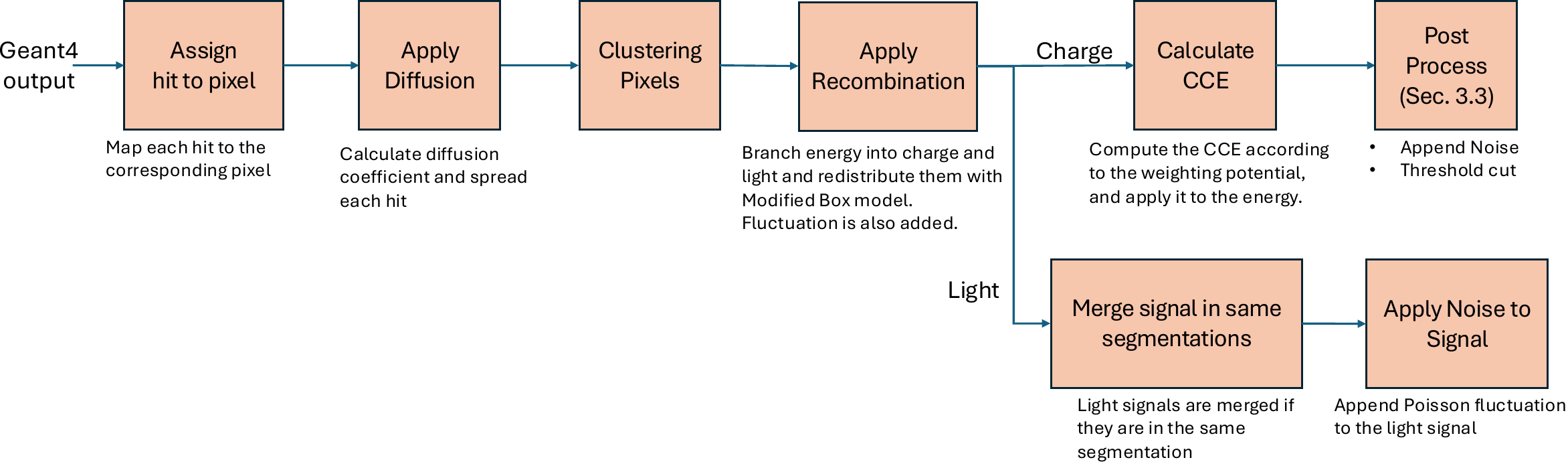}
		\caption{Workflow of the simulation}\label{fig:workflow_simulation}
	\end{figure*}

	Figure~\ref{fig:workflow_simulation} shows the workflow of the simulation. After Geant4 generated the hit information for each interaction, each hit was assigned to the corresponding pixel, the diffusion process (Section \ref{subsubsec:diffusion}) was applied, and hits spreading over multiple pixels were clustered. Then electron-ion recombination was applied, splitting the energy deposition into light and charge signals. For the charge signal, the charge collection efficiency was calculated, and the noise and threshold were then applied in the post-processing (Section \ref{subsec:postprocess}), while light signals within the same segment were merged and Poisson fluctuation was appended.
	In the real situation, the recombination process and the effect of the charge collection efficiency take effect before the diffusion process. However, in this simulation, we adopted this order because the calculation of the recombination uses the total energy deposition of the hits as the kinetic energies of the secondary electrons (see Section \ref{subsubsec:recombination}).

	\begin{figure}
		\centering
		\includegraphics[keepaspectratio, scale=0.3, page=1]{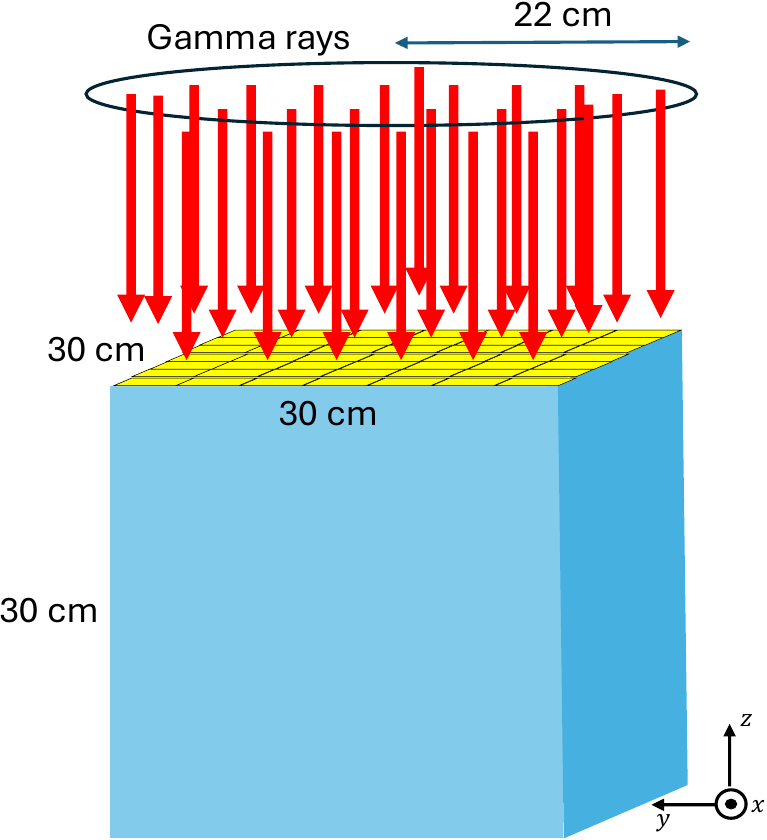}
		\caption{Geometry in the Geant4 simulation}\label{fig:geometry}
	\end{figure}
	\subsection{Geant4 simulation}\label{subsec:geant4}
		We considered a \qtyproduct{30 x 30 x 30}{\centi\meter} liquid-argon box with no additional passive materials in the simulation world. We assume that, in the actual detector, the segments, described in Section \ref{sec:detector}, are separated by walls. In this simulation, however, they are implemented as \qtyproduct{2.5 x 2.5 x 30}{\centi\meter} logical volumes only and the walls are not modeled as passive material for simplicity. Figure~\ref{fig:geometry} shows the simulation geometry.

		We defined a pixelated anode with pixel pitches of \qtylist{0.5;1;3.125}{mm}, and the $z$ coordinate was smeared by a Gaussian with $\sigma = \qty{0.26}{mm}$ independent of the pixel pitch and the $z$ coordinate. This value was chosen as the typical value of the longitudinal diffusion (see Section \ref{subsubsec:diffusion}).
		We irradiated the detector with monochromatic photons of various energies (\qtylist{511;700;900;1173.2;1332.5;2000}{\keV}) from above the detector toward the anode.
		The energy deposition was integrated along the $z$-axis for each pixel. The shape of the beam was a circle of radius \qty{22}{cm} to cover the entire detector.
		We used the Livermore model as the electromagnetic model for the Geant4 simulation, which considers the Doppler broadening of the bound electrons in argon.

	\subsection{Processes in LAr}\label{sec:simulation:processes}
		In liquid argon, several effects can impact the energy and position resolutions. We considered four effects and implemented them in the simulation.

		In this simulation, we assumed ideal conditions for simplicity; therefore, certain factors were not implemented. We assumed a uniform electric field without distortions due to instrumentation or space charge effects. We also assumed a sufficiently high argon purity such that the electron lifetime is much longer than the maximum drift time (\qty{\sim 180}{\micro\second} in this configuration) in the detector.

		\subsubsection{Diffusion}\label{subsubsec:diffusion}
			While ionization electrons drift, they spread due to thermal motion, which may affect the position resolution of each hit.
			For an electric field strength of \qty{500}{\volt\per\cm}, we used transverse and longitudinal diffusion coefficients of \qty{12.0}{\square\cm\per\second} and \qty{7.2}{\square\cm\per\second}, respectively \citep{diffusion}.

			In the simulation, the electron cloud was divided into 64 particles to implement the transverse diffusion. For each particle, the position was sampled from a normal distribution, after deriving the width of the distribution from the diffusion coefficient $D$, drift velocity $v$ and the $z$ coordinates with $\sigma=\sqrt{2Dz/v}$. For the longitudinal diffusion, we used $\sigma = \qty{0.26}{\mm}$ in the $z$ coordinate as a typical value, which was calculated by the same formula and taken as half of the value at the maximum drift distance. Also, we used the drift velocity of \qty{1.65}{\milli\meter\per\micro\second}, calculated from \citep{diffusion}.

			After diffusion was applied, for signals spread across multiple pixels, we performed the clustering process. We merged adjacent detected pixels. In the merging process, the position is weighted by the measured energy.

		\subsubsection{Recombination}\label{subsubsec:recombination}
			After ionization electrons are produced by energy deposition, some of them recombine with argon ions. This process reduces the number of drifting electrons; as a result, the measured energy deposition would be underestimated when using only the charge signal.

			We used the Modified Box model \citep{ModifiedBox} to account for recombination. In this model, the ratio, $R$, of electrons surviving after recombination to the initial number of electrons is expressed as:
			\begin{equation}\label{eq:modifiedBox}
				R\left(\dedxinline\right) =  \dfrac{\log\left(\alpha + \dedx \frac{\beta}{\epsilon\rho}\right)}{\dedx\frac{\beta}{\epsilon\rho}},
			\end{equation}
			where $\alpha=0.92$, $\beta=\qty{0.184}{\kilo\volt\gram\per\cubic\centi\meter\per\mega\electronvolt}$ \citep{MicroBooNECalib2020}, $\rho=\qty{1.396}{\gram\per\cubic\centi\meter}$ is the density of LAr, and $\epsilon$ is the electric field (\qty{500}{\volt\per\cm} in this simulation).

			We also calculated the number of scintillation photons to use for the energy correction for recombination. We define quanta $N_{q}$ as the total number of excitations (photon) $\nph$ and ionizations (electron) $\nel$:
			\begin{equation}
				\nq = \nph + \nel.
			\end{equation}
			Here, the number of quanta and the number of surviving electrons are expressed as:
			\begin{align}
				\nq  & = \frac{E}{W_\mathrm{q}},                            \\
				\nel & = R\left(\dedxinline\right)\frac{E}{W_\mathrm{ion}},
			\end{align}
			where $E$ is the energy deposition, and $W_\mathrm{q}, W_\mathrm{ion}$ are the W-values for quanta and ionization in LAr, respectively. We adopted $W_\mathrm{q}=\qty{19.5}{eV}$ \citep{2002JaJAP..41.1538D} and $W_\mathrm{ion} = \qty{23.6}{eV}$ \citep{1974PhRvA...9.1438M}.

			Therefore, the number of scintillation photons is represented as:
			\begin{equation}\label{eq:scintillation}
				\nph = \nq - \nel.
			\end{equation}
			The charge yield was assumed to fluctuate according to a binomial distribution (denoted as $\mathrm{Binom}(N, r)$) and a Gaussian fluctuation of the escaping probability of the electrons. This model is based on that of liquid xenon (LXe) \citep{2017PhRvD..95a2008A}. It has one parameter $\sigma_p$, but no experimental measurements exist for LAr. Therefore, we used the same value in LXe (0.06 in \citep{2017PhRvD..95a2008A}).

			We denote a fluctuated value as the value with a prime, hereafter.
			Firstly, we sampled the probability from a Gaussian distribution with a standard deviation of $\sigma_p$.
			\begin{equation}\label{eq:prob_fluc}
				R' = \mathrm{Gauss}\left(R, \sigma_p \right).
			\end{equation}
			We sampled the number of electrons from a binomial distribution:
			\begin{equation}\label{eq:charge_fluc}
				\nel' = \mathrm{Binom}\left(\nel, R'\right).
			\end{equation}
			The number of scintillation photons is calculated with Eq.~\ref{eq:scintillation}.

			As described in Section~\ref{sec:detector}, the segmentation allows the light from individual hits to be measured separately. The light collection efficiency (LCE) should depend on the position of the energy deposition, but it is set to \qty{10}{\percent} as the standard value, independent of the position for simplicity. Its dependence is discussed in Section~\ref{sec:ResultsAndDiscussion}.

			We implemented the recombination model in \comptonsoft\  as follows:
			\begin{enumerate}
				\item We consider the kinetic energy of the secondary electron as the total energy deposition at a hit. The total energy deposition is taken as that of the clustered hits.
				\item Calculate $\dedxinline$ from the kinetic energy by using the ESTAR database \citep{ESTAR2005}.
				\item Calculate the charge yield by Eq.~\ref{eq:charge_fluc}.
				\item The number of scintillation photons is calculated by Eq.~\ref{eq:scintillation} and multiplied by the light collection efficiency.
			\end{enumerate}

		\subsubsection{Charge Collection Efficiency (CCE)}\label{subsubsec:cce}
			We considered the effect of the induced signal on the anode.
			In LAr, the drift velocity of argon ions is much lower than that of electrons \citep{ArgonionMobility}, and the charge signal is almost entirely due to electron drift, which reduces the signal especially near the anode.  According to the Shockley-Ramo theorem \citep{ShockleyRamo}, the induced charge on the anode is calculated as follows:
			\begin{equation}
				C_i = C_\mathrm{ion}\left(\phi_{i}(z) - \phi_i(z_{\mathrm{anode}})\right),
			\end{equation}
			where $C_i$ is the induced charge on pixel $i$, $C_\mathrm{ion}$ is the charge of the ionized electrons remaining after recombination is applied, and $\phi_{i}(z)$ is the weighting potential at position $z$ (not smeared by the Gaussian distribution described in Section \ref{subsec:geant4}), defined as the electric potential distribution under the boundary condition that pixel $i$ is set to unity and all other electrodes to zero. We calculated the weighting potential using the analytic solution of pixel detectors described in \citep{ODAKA2010303}.

		\subsubsection{Energy Resolution Due to Statistical Fluctuation and Electronic Readout Noise}\label{subsubsec:energy_resolution}
			We also applied energy fluctuation to the light/charge signal caused by statistics and the charge readout system.

			For the charge signal, we applied the noise formulated in Eq.~\ref{eq:noise_charge}:
			\begin{equation} \label{eq:noise_charge}
				\sigma^2 = p_0^2 + p_1^2 E.
			\end{equation}
			The first and second terms correspond to electronic and statistical noise, respectively. We fixed $p_1$ to \qty{0.05}{\keV^{0.5}}, which includes the Fano fluctuation with a Fano factor of 0.107 \citep{1976NucIM.134..353D}.

			For the light signal, we applied Poisson statistics of photon counts detected by a SiPM.
	\subsection{Post-processing of the simulated signals}\label{subsec:postprocess}
		After the recombination and CCE were applied, we appended some effects to the charge signal.
		We applied the noise to the charge signals as described in Section \ref{subsubsec:energy_resolution} and removed the hits with an energy below the threshold (5 times the electronic noise). It may be more realistic to apply the charge noise and the threshold before clustering. However, we applied these effects in order to isolate the pure effect of the energy resolution affecting the detector response.

\section{Data Processing}\label{sec:data_process}
	\subsection{Energy Correction}\label{subsec:energyCorrection}
		Since the energy deposition is reduced by the processes described above, we applied an energy correction to the simulation results in 2 steps: CCE and recombination corrections.

		As we explained in Section \ref{subsubsec:cce}, the signals decrease if the energy deposition occurs near the anode. To avoid this effect, we cut the hits whose $z$ coordinate is above a threshold. We took \qtylist{139;144;117}{\mm} from the center of the detector as the thresholds for the pixel pitches of \qtylist{1;0.5;3.125}{\mm}, respectively.

\begin{figure}
			\centering
			\includegraphics[keepaspectratio, width=0.9\linewidth]{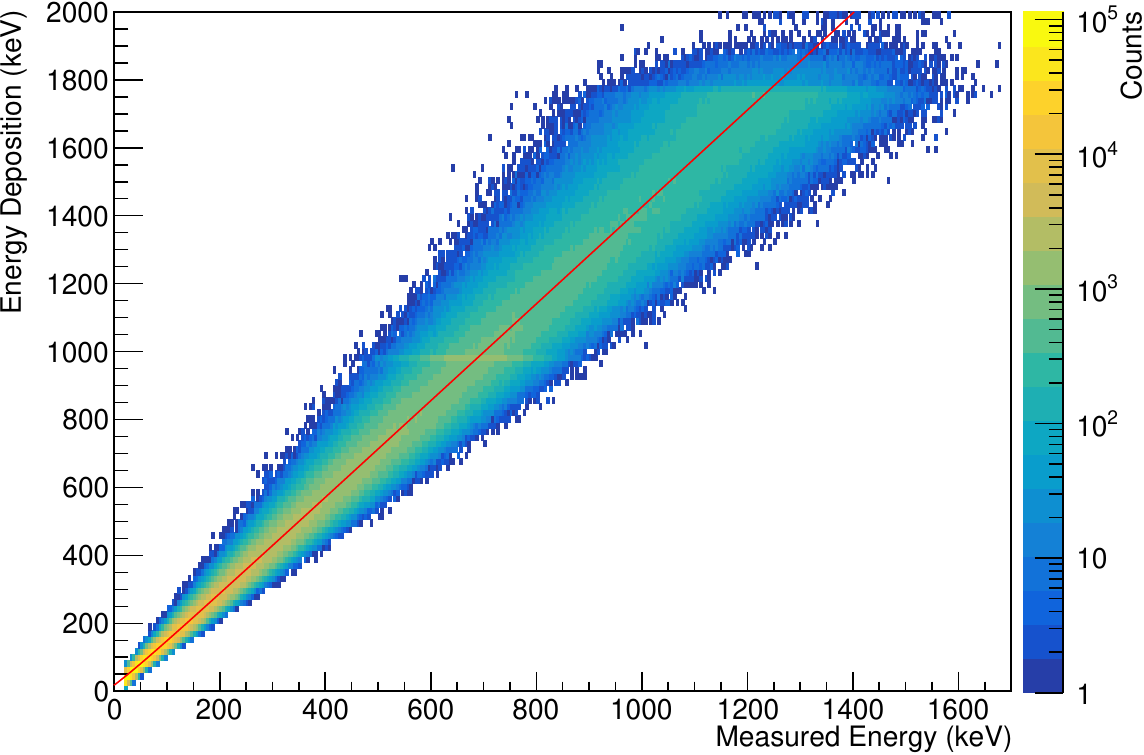}
			\caption{Fitted gain function and the relationship between the energy deposition and the measured energy. The red line shows the fitted function.}\label{fig:gain_function}
		\end{figure}

		As the second step, we corrected for the energy loss due to recombination using three different methods to study the effect of using the light signal.
		First of all, we used the light signal as well as the charge signal with Eq.~\ref{eq:scintillation}. Since the light signals are merged in a segment, we selected events in which all hits belonged to different segments.
		The second way is to use the charge signal only. We shot a \qty{2}{\MeV} photon beam horizontally and created a 2D histogram showing the measured energy deposition (with all detector effects applied) versus the true deposited energy (without any detector effects), as shown in Figure~\ref{fig:gain_function}. As the first step, we performed a Gaussian fit to the measured energy distribution in each bin of the energy deposition to derive representative points in the 2D histogram. After that, we fitted these points with this formula:

		\begin{equation}\label{eq:fitting_function}
			E_\mathrm{dep, measured} = \frac{p_0E_{\mathrm{dep, true}}^2 + p_1E_{\mathrm{dep, true}} + p_2}{p_3E_{\mathrm{dep, true}} + p_4}.
		\end{equation}
		In the fitting procedure, we ignored the bin including \qty{978}{keV} because this energy has a peak in the spectrum due to the pair production of the incident photon. We derived the inverse function analytically to apply this function to the simulation.
		Finally, we used a combination of the previous methods. We count the number of hits in one segment, and use the light signal if all hits are in different segments, otherwise we use the correction function derived from the 2D histogram. Note that the segmentation condition is used here only to select which correction to apply, and no events are rejected. We call this method the adaptive correction hereafter.

	\subsection{Event Reconstruction} \label{subsec:event_reconstruction}
		The event reconstruction proceeds in two steps.
		As the first step, the hit order is determined by sorting the interaction time recorded in the Geant4 simulation. This approach relies on the time information of each interaction, which is not directly available from an actual detector. However, the aim of this study is to clarify how physical effects in liquid argon degrade the event-reconstruction performance rather than to evaluate specific reconstruction algorithms. From this perspective, we adopted this method.

\begin{figure}
			\centering
			\includegraphics[page=2, keepaspectratio, width=0.8\linewidth]{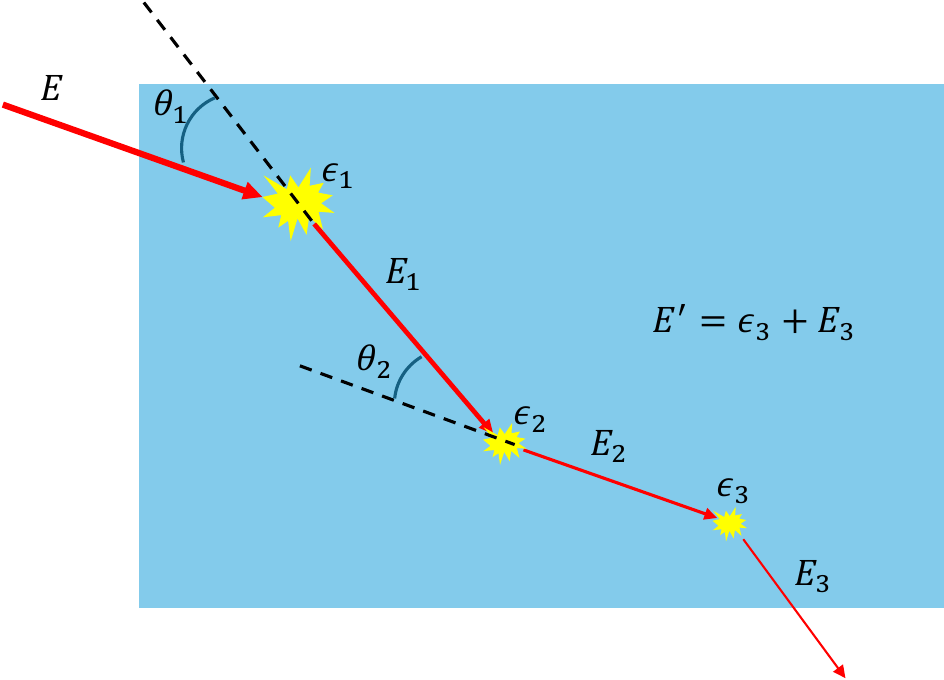}
			\caption{Notation for Eq.~\ref{eq:event_reconstruction_e}, \ref{eq:event_reconstruction_cos}, \ref{eq:event_reconstruction_e2}}\label{fig:event_reco}
		\end{figure}

		As the next step, we selected events with at least 3 hits, and reconstructed them by the following formulae:
		\begin{align}
			E            & = \epsilon_1 + \epsilon_2 + E_2, \label{eq:event_reconstruction_e}                                 \\
			\cos\theta_1 & = 1 - m_ec^2\left(\frac{1}{E-\epsilon_1} - \frac{1}{E}\right), \label{eq:event_reconstruction_cos}
		\end{align}

		\begin{equation}
			E_2 = \left\{
			\begin{array}{ll}
				\epsilon_3                                                                                          & \text{3rd hit is PA} \\
				\begin{array}[t]{l}
					-\dfrac{\epsilon_2}{2}+\sqrt{\dfrac{\epsilon^2_2}{4} + \dfrac{\epsilon_2 m_e c^2}{1-\cos\theta_2}}
				\end{array} & \text{otherwise}
			\end{array}
			\right.
			\label{eq:event_reconstruction_e2}
		\end{equation}
		where $\epsilon_i, \theta_i$ denote the energy deposition and the scattering angle at the $i$-th scattering point, $E_i$ is the energy of the gamma ray after the $i$-th scattering, $E$ is the energy of the incident gamma ray (Figure~\ref{fig:event_reco}), and PA means photoabsorption.
		Similarly to the first step, the classification of each interaction as photoabsorption or Compton scattering, used in Eq.~\ref{eq:event_reconstruction_e2}, was based on the process recorded in the Geant4 simulation.
		The fractions of events with more than 3 hits, 3-hit events with photoabsorption at the final interaction, and 3-hit events without photoabsorption were \qty{46.8}{\percent}, \qty{47.3}{\percent}, and \qty{5.9}{\percent}, respectively, for the condition with a noise of \qty{4}{keV}, a pixel pitch of \qty{1}{\mm}, the light and charge recombination correction and a \qty{1332.5}{\keV} incident photon.

\section{Results and Discussion}\label{sec:ResultsAndDiscussion}
	\begin{figure*}[htb]
		\begin{subfigure}{0.48\textwidth}
			\centering
			\includegraphics[width=0.8\linewidth]{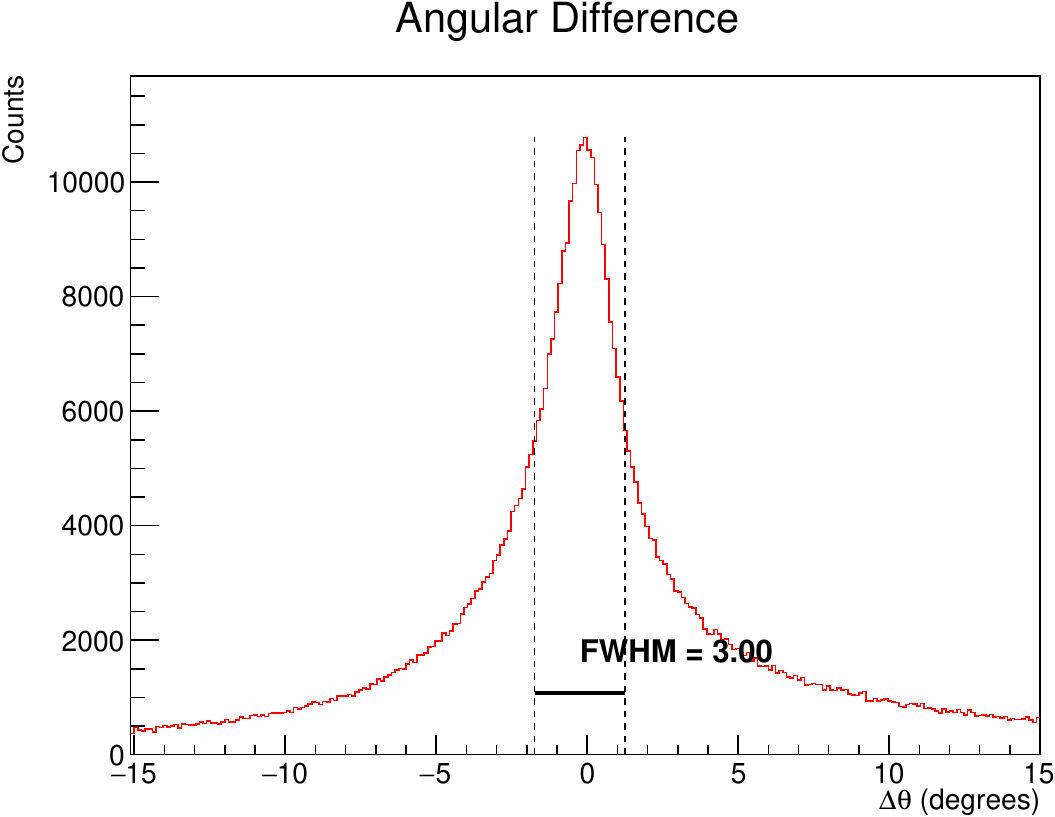}
		\end{subfigure}
		\begin{subfigure}{0.48\textwidth}
			\centering
			\includegraphics[width=0.8\linewidth]{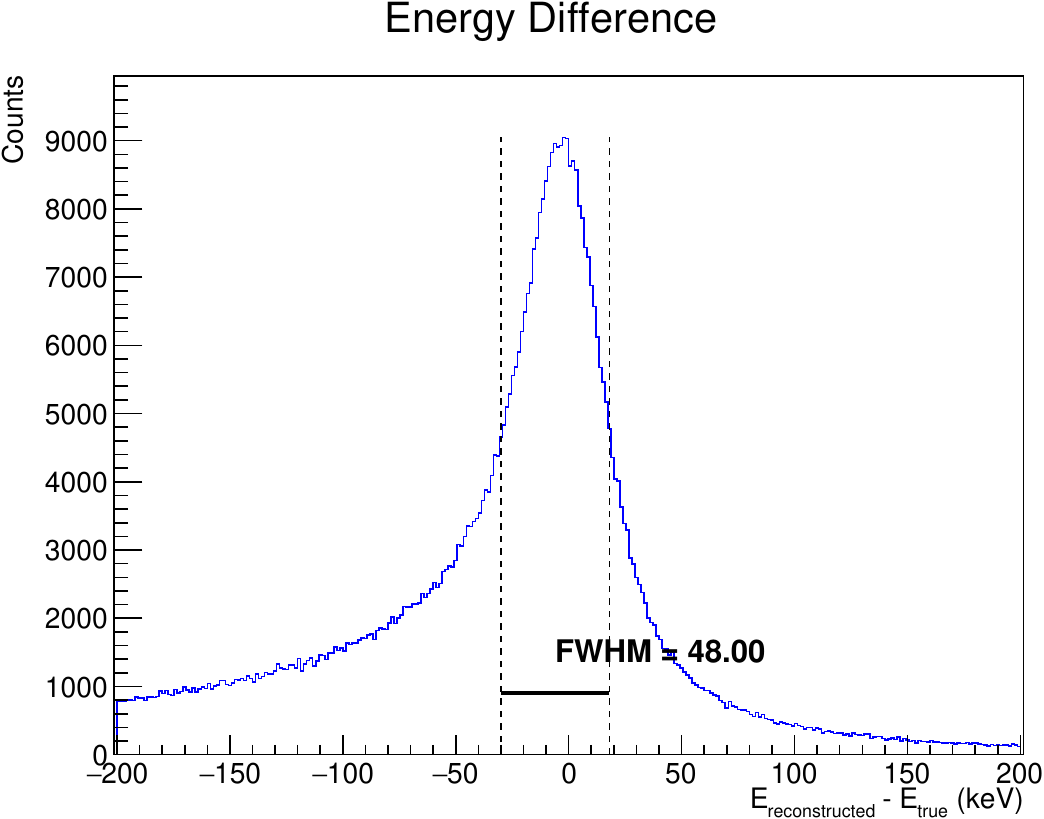}
		\end{subfigure}
		\caption{An example of the ARM and reconstructed energy error distributions with a \qty{1332.5}{\keV} incident photon, a pixel pitch of \qty{1}{\mm}, a \qty{4}{\keV} energy resolution, and the light and charge correction. The FWHMs calculated from the histograms are also shown.}\label{fig:example}
	\end{figure*}

	\begin{figure*}[tb]
		\begin{subfigure}{0.48\textwidth}
			\centering
			\includegraphics[width=0.8\linewidth]{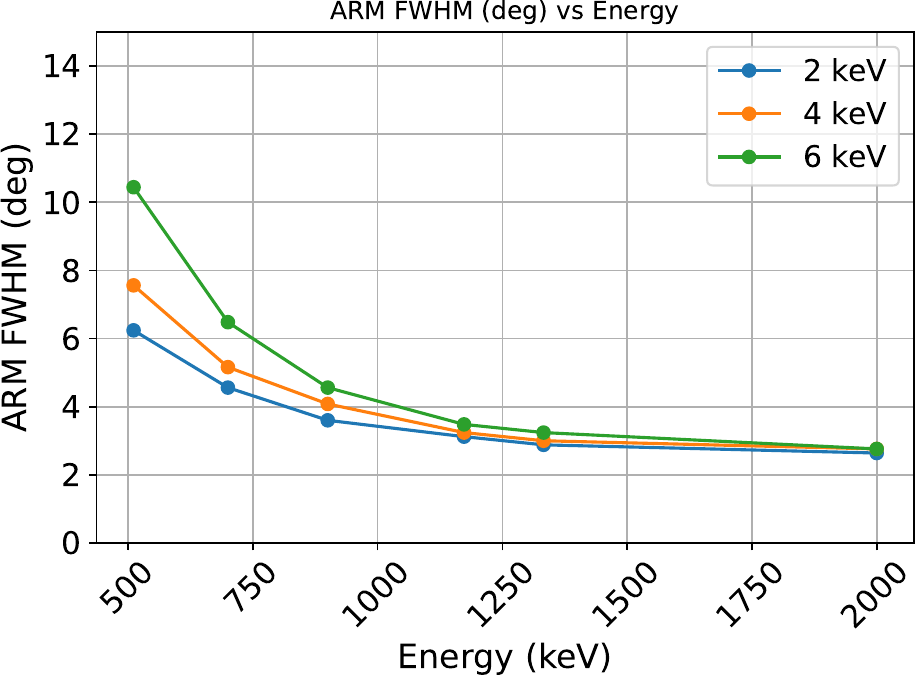}
		\end{subfigure}
		\begin{subfigure}{0.48\textwidth}
			\centering
			\includegraphics[width=0.8\linewidth]{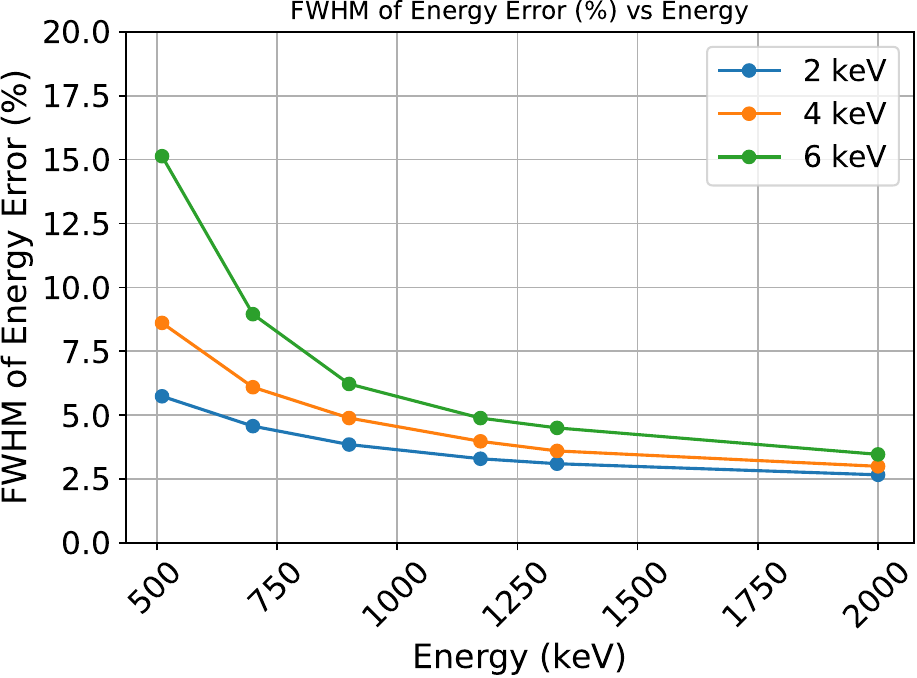}
		\end{subfigure}
		\caption{The dependence of the detector performance on the noise level.
			The simulation was performed under the condition of a \qty{1}{\mm} pixel pitch with the charge + light recombination correction.
			The legend indicates the different values of the electronic noise, and the energy resolution also includes the statistical term with applying the recombination effect. The errors of the angular resolution and the energy resolution are typically \qty{\sim0.2}{\degree} and \qty{\sim 0.1}{\percent}, respectively, and are smaller than the marker size.}\label{fig:noise}
	\end{figure*}
	\begin{figure*}[tb]
		\begin{subfigure}{0.48\textwidth}
			\centering
			\includegraphics[width=0.8\linewidth]{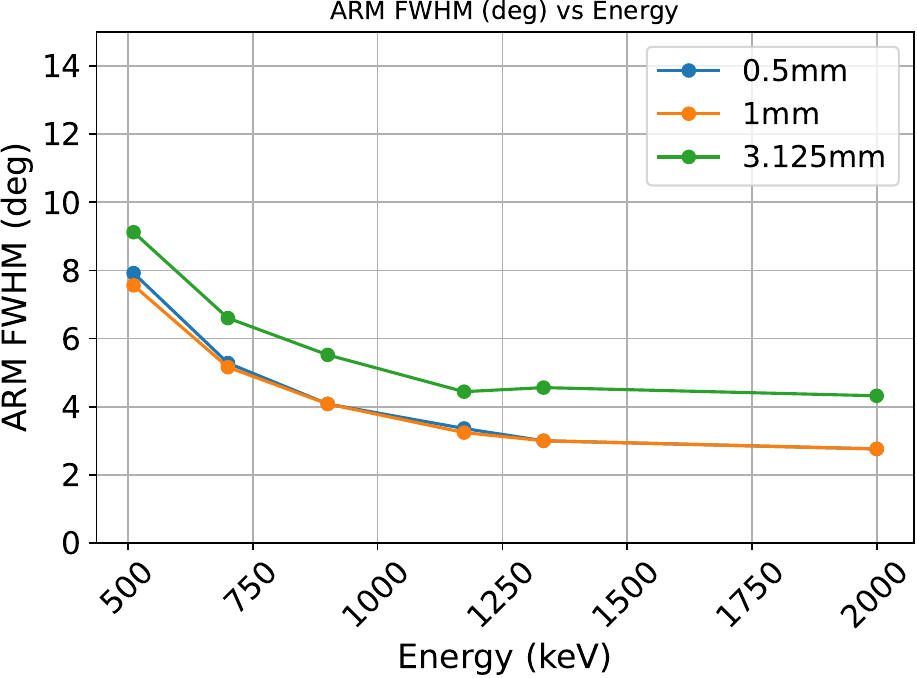}
		\end{subfigure}
		\begin{subfigure}{0.48\textwidth}
			\centering
			\includegraphics[width=0.8\linewidth]{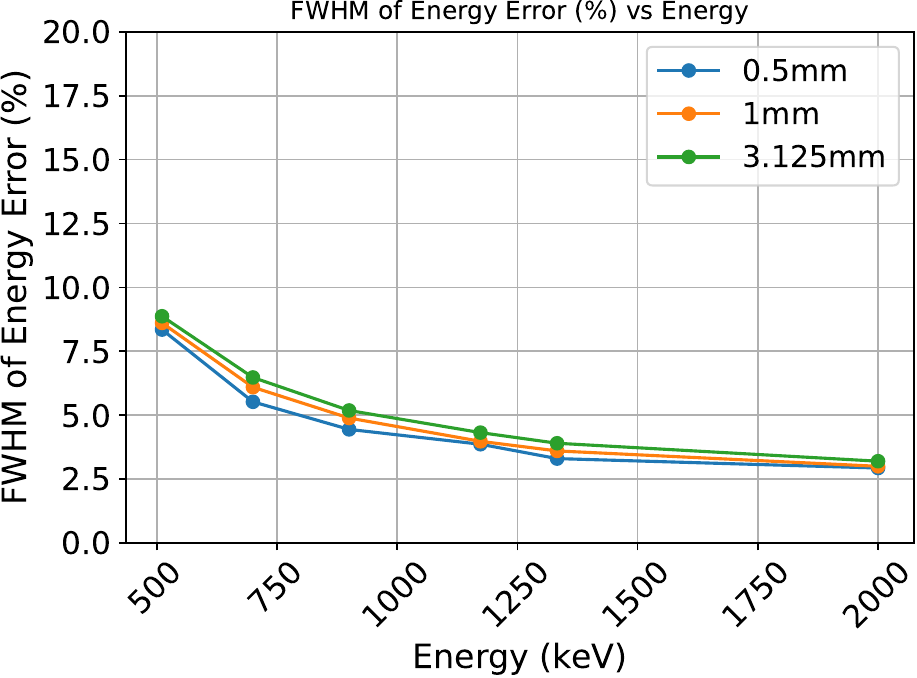}
		\end{subfigure}
		\caption{Same as Figure~\ref{fig:noise}, but for the dependence on the pixel pitch.}\label{fig:pitch}
	\end{figure*}

	In this study, we evaluated the performance of the detector using the ARM and the error of the reconstructed energy. Figure~\ref{fig:example} shows an example of the ARM distribution and the energy error ($E_\mathrm{reconstructed} - E_\mathrm{true}$) in the event reconstruction. The FWHM (full width at half maximum) for each value is used for the evaluation.

	Here, we define the standard condition of this study: the standard deviation of the electronic noise is \qty{4}{\kilo\electronvolt}, the pixel pitch is \qty{1}{\mm} and the correction method is the light + charge method with \qty{10}{\percent} of the light collection efficiency (LCE). From this condition, we changed four parameters: the energy resolution, the pixel pitch, the energy correction method and LCE.
	\begin{figure*}[tbh]
		\begin{subfigure}{0.48\textwidth}
			\centering
			\includegraphics[width=0.8\linewidth]{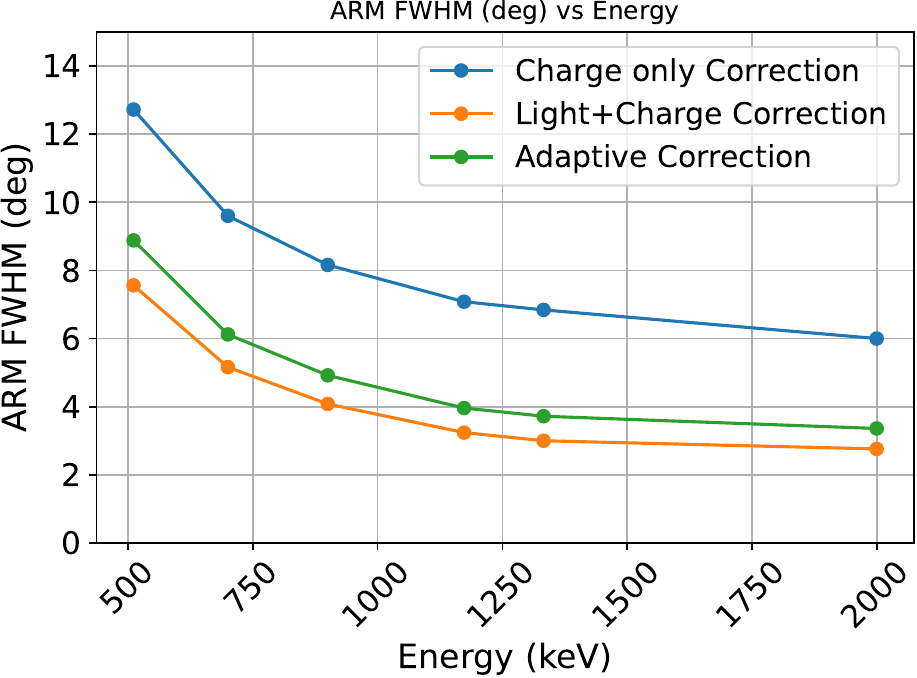}
		\end{subfigure}
		\begin{subfigure}{0.48\textwidth}
			\centering
			\includegraphics[width=0.8\linewidth]{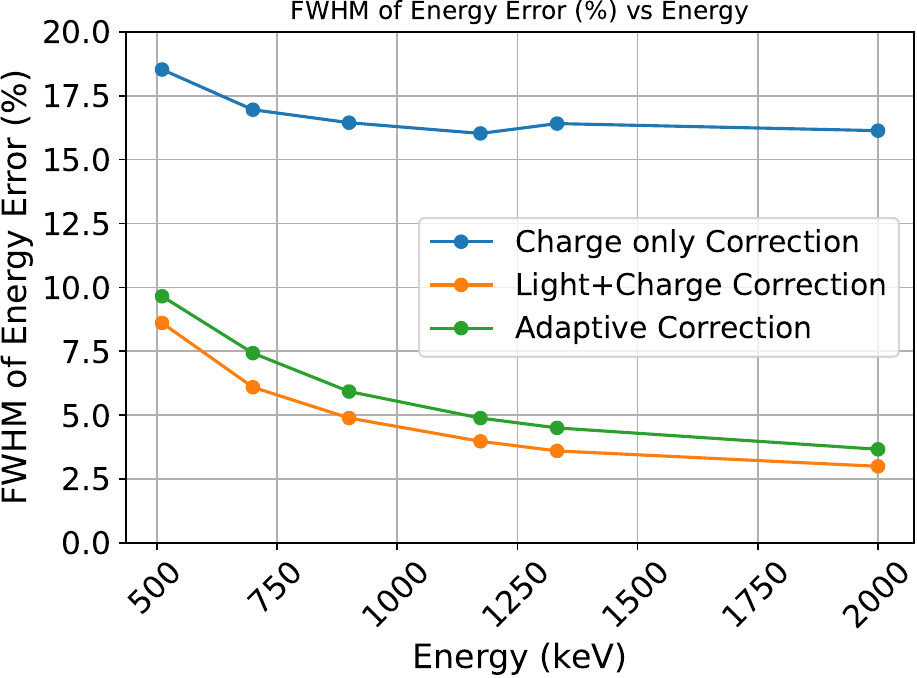}
		\end{subfigure}
		\caption{Same as Figure~\ref{fig:noise}, but for the dependence on the recombination correction method.}
		\label{fig:correction}
	\end{figure*}

	Figure~\ref{fig:noise} shows the dependence of the ARM FWHM and the error of the reconstructed energy on the readout electronic noise.  At \qty{1332.5}{\keV}, a noise of \qty{4}{\keV} yields an ARM of about \qty{3}{\degree}.
	This figure shows that the noise level has little effect on the angular resolution at higher photon energies.
	The angular resolution affected by the energy resolution, $\Delta\theta_{E}$, is approximately proportional to $\sigma_E/E$ \citep{GRAMS2020};
	therefore, the higher $E$ is, the smaller $\Delta\theta_{E}$ becomes.
	This ARM is almost the same as that of the Compton Spectrometer and Imager (COSI), even though COSI uses a germanium detector with higher energy resolution: the ARM for \qty{1332.5}{\kilo\electronvolt} gamma rays is \qty{4.0}{\degree} \citep{2022NIMPA103166510B}. This can be explained by the distance between the first 2 hits. The mean free path of \qty{1}{\mega\electronvolt} photons is \qty{12.6}{\centi\meter} in LAr, compared with \qty{3.3}{\centi\meter} in germanium. Therefore, as long as the TPC is large enough, the distance between the first two interactions tends to become longer, which improves the geometrical contribution to the angular resolution and compensates for the poorer energy resolution.

	For the energy error, Figure~\ref{fig:noise} shows that a noise of \qty{4}{\keV} achieves an error of less than \qty{4}{\percent} above \qty{1173.2}{\kilo\electronvolt}, which is sufficient for astrophysical observations. The \qty{6}{\keV} condition gives relatively worse but still acceptable results in the same energy range.

	Figure~\ref{fig:pitch} shows how the pixel pitch affects the detector performance.
	This figure shows that the position resolution has an impact primarily on the ARM. The \qty{1}{\mm} pitch yields approximately \qtyrange[range-phrase={--}]{64}{87}{\percent} of that obtained with the \qty{3.125}{\mm} pitch while it gives almost the same ARM as that of the \qty{0.5}{\mm} pitch.
	This behavior can be understood by comparing the contributions to the position resolution of a single hit.
	The pixel pitch contributes $\sigma_{xy} = \qtylist{0.90;0.29;0.14}{\milli\meter}$ for pixel pitches of \qtylist{3.125;1;0.5}{\milli\meter}, respectively. The other contributions are independent of the pitch: the smearing of the $z$ coordinate, $\sigma_z \sim \qty{0.26}{\milli\meter}$; the transverse diffusion, $\sigma_\mathrm{diffusion} \sim \qty{0.33}{\milli\meter}$ (half of the diffusion at the maximum drift length); and the spread of \qty{1.0}{\mega\electronvolt} electrons $\sigma_{e} \sim \qty{1.1}{\milli\meter}$ (obtained from Geant4 simulation). At the \qty{3.125}{\milli\meter} pitch, $\sigma_{xy}$ becomes comparable to the spread of the electrons and therefore worsens the ARM, whereas at \qty{1}{\milli\meter} it is already smaller than $\sigma_{e}$ and $\sigma_\mathrm{diffusion}$. Reducing the pitch further to \qty{0.5}{\milli\meter} thus brings no significant improvement.

	In addition, Figure~\ref{fig:pitch} suggests that the difference in the energy error among the different pixel pitches is small.
	This is because our simulation applies the noise after the clustering process and the position resolution affects only the energy estimate of the escaping photon beyond the third interaction (note that our reconstruction method uses only the first 3 hits).
	Considering the technical difficulties of achieving \qty{0.5}{\mm} pitch pixels with high energy resolution and a low complexity of the circuit, the \qty{1}{\mm} pitch is the best option unless electron tracking is considered.

	\begin{table*}[htbp]
		\centering
		\caption{Step-by-step event cut rate at \qty{1332.5}{\keV}. The values show the fraction of events surviving each selection step.}
		\label{tab:cut_rate_merged}
		\begin{threeparttable}[htbp]
			\begin{tabular}{l r r r r r}
				\hline
				\hline
				Method       & \multicolumn{4}{c}{Surviving fraction at each step}                                                                           & Final event efficiency \\
				             & Detection\tnote{1}    & Segmentation filter\tnote{2} & Event reconstruction\tnote{3} & ARM filter within \qty{\pm12}{\degree} &                        \\
				\hline
				Light+Charge & \qty{81.38}{\percent} & \qty{84.16}{\percent}        & \qty{18.33}{\percent}         & \qty{70.15}{\percent}                  & \qty{8.81}{\percent}   \\
				Charge Only  & \qty{81.38}{\percent} & \qty{100.00}{\percent}       & \qty{26.04}{\percent}         & \qty{68.28}{\percent}                  & \qty{14.47}{\percent}  \\
				Adaptive     & \qty{81.38}{\percent} & \qty{100.00}{\percent}       & \qty{24.84}{\percent}         & \qty{70.93}{\percent}                  & \qty{14.34}{\percent}  \\
				\hline
			\end{tabular}
			\begin{tablenotes}[para,flushleft,normal] 
				\item[1] The ratio between the number of events where any charge signals are detected and the number of incident photons (the number of photons multiplied by the geometrical effect).
				\item[2] This cut requires that each segment contains no more than one hit. It is applied only for the light + charge correction; no events are rejected for the other two methods (expressed as \qty{100}{\percent}).
				\item[3] This step requires at least three hits after removing the hits near the anode for the CCE correction (Section \ref{subsec:energyCorrection}), and rejects non-physical reconstruction results (e.g., the cosine of the scattering angle falling outside the range from $-1$ to $1$).
			\end{tablenotes}
		\end{threeparttable}
	\end{table*}

	In addition to the energy and the position errors, the detector performance should depend on the recombination correction method. The difference among the three methods described in Section~\ref{subsec:energyCorrection} is shown in Figure~\ref{fig:correction}.
	The correction with light and charge gives a 1.7--2.3 times better ARM than the charge-only correction, and the former provides a 2.2--5.4 times better energy error than the latter. The adaptive correction yields intermediate results.
	However, the stricter event selection reduces the effective area. We calculated the efficiency as the ratio of the events whose ARM was within \qty{\pm 12}{\degree} (wide enough to cover the main ARM peak at \qty{1332.5}{\kilo\electronvolt}) to the incident events.

	Table~\ref{tab:cut_rate_merged} shows the step-by-step event cut rates for the three recombination correction methods at \qty{1332.5}{\keV}. The final event efficiencies are \qty{8.8}{\percent}, \qty{14.5}{\percent}, and \qty{14.3}{\percent}, for the light and charge correction, the charge-only correction, and the adaptive correction, respectively, which correspond to effective areas of \qty{79.3}{\square\centi\meter}, \qty{130.2}{\square\centi\meter}, and \qty{129.1}{\square\centi\meter}.
	Since we use the same Geant4 simulation data, the numbers of detected events are the same. The event reconstruction step retains fewer events for the light + charge correction (\qty{18.3}{\percent}) than for the other two methods (\qty{\sim25}{\percent}). This is because the segmentation cut preferentially removes multiple-hit events, in which two interactions are more likely to fall into the same segment, and thus biases the surviving sample toward a lower hit multiplicity. The subsequent requirement of at least three hits then rejects a larger fraction of these events.
	The adaptive correction avoids this loss because, as described in Section~\ref{subsec:energyCorrection}, the segmentation condition is used only to select which correction to apply, and no events are rejected by it.

	As a result, the light + charge correction reduces the efficiency to \qty{61}{\percent} of that of the charge-only correction. On the other hand, the combination of these methods gives almost the same efficiency as the charge-only correction while limiting the degradation of the ARM and the energy error by \qty{\sim20}{\percent} compared to the light + charge correction.
	Since the	charge-only correction relies on Monte Carlo information to derive the correction function, its performance shown in Figure~\ref{fig:correction} is itself an optimistic estimate. Our results therefore indicate that the use of both light and charge signals is necessary to achieve better performance.

\begin{figure*}[tb]
		\begin{subfigure}{0.48\textwidth}
			\centering
			\includegraphics[scale=0.5]{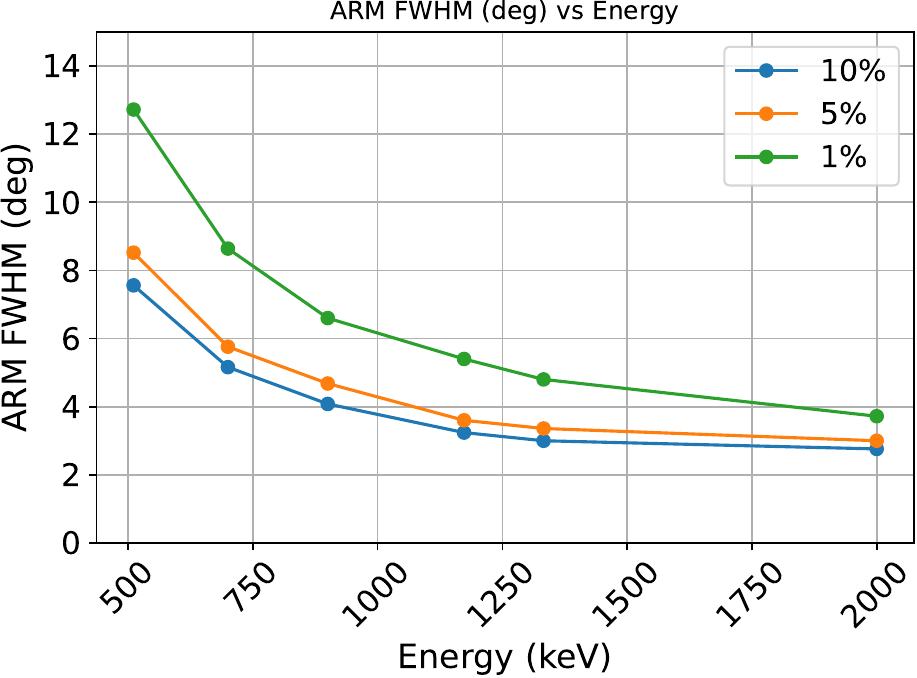}
		\end{subfigure}
		\begin{subfigure}{0.48\textwidth}
			\centering
			\includegraphics[scale=0.5]{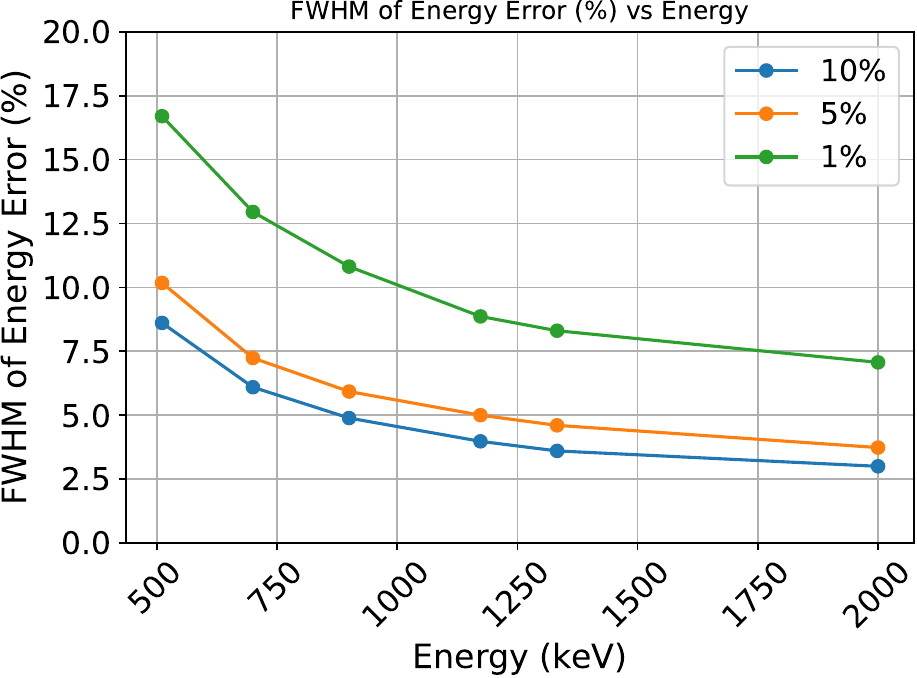}
		\end{subfigure}
		\caption{Same as Figure~\ref{fig:noise}, but for the dependence on the light collection efficiency.}\label{fig:lce}
	\end{figure*}

	We also investigated how much impact the LCE has on the detector performance, for LCE values of \qtylist{10;5;1}{\percent}. Figure~\ref{fig:lce} shows the LCE dependences of the detector performance. These figures show that the performance difference between LCEs of \qty{1}{\percent} and \qty{5}{\percent} is larger than that between \qty{5}{\percent} and \qty{10}{\percent}. Even with \qty{5}{\percent} LCE, the FWHMs of ARM and the energy difference are \qty{\sim10}{\percent} and \qty{\sim20}{\percent} worse than those with a \qty{10}{\percent} LCE, so a \qty{5}{\percent} LCE would also be acceptable if a \qty{\sim10}{\percent} LCE turns out to be technically difficult. Considering that some LArTPC experiments achieve a light collection efficiency of \qty{10}{\percent} \citep{2002JaJAP..41.1538D,2021EPJC...81.1014A}, we adopt \qty{10}{\percent} as a design target for future Compton cameras.

	We here discuss the feasibility of the assumed conditions and possible uncertainties of the simulation in this work.
	First, an electronic noise of \qty{4}{\keV} can be achieved, since NanoGRAMS has an equivalent noise charge (ENC) of $130$--$160 e^-$ with a \qty{3.2}{\milli\meter} pitch \citep{takashima}, which corresponds to \qtyrange[range-phrase={--}]{3.1}{3.8}{\keV} since $W_\mathrm{ion} = \qty{23.6}{eV}$. A smaller pixel pitch generally gives a smaller detector capacitance, and lower electronic noise.

	Second, the recombination model still has uncertainties.
	Therefore, for Compton camera development, it is essential to perform a precise measurement of this effect using a real instrument. In particular, the recombination fluctuation has not been studied in detail, while there are some studies on liquid xenon \citep{2017PhRvD..95a2008A,2025FrDST...280975S}.

	Finally, in a more realistic situation, where the pixel pitch affects the noise performance, the hit multiplicity in the clustering process may also affect the energy resolution because the total energy resolution accumulates during the clustering. In this paper, we applied the noise after the clustering process to investigate the pure influence of the energy resolution.  We should optimize the electrode configuration such as the pitch and the shape (pixel or strip), considering the energy resolution in the future.

\section{Conclusion}\label{sec:conclusion}
	We conducted a simulation-based performance study of Compton cameras using LArTPC detector technology. We considered electron-ion recombination, the charge induction on the anode, and charge diffusion.
	We found that a noise level of \qty{4}{\keV} and a pixel pitch of \qty{1}{\mm} yield an ARM of about \qty{3}{\degree} at \qty{1332.5}{\kilo\electronvolt} and an energy reconstruction error better than \qty{4}{\percent} above \qty{1173.2}{\keV}. The recombination correction using both light and charge signals gives a significant improvement over the charge-only correction. However, the former reduces the efficiency to about \qty{60}{\percent} of that of the latter, and the use of the adaptive correction maintains high performance while suppressing the efficiency loss. These numbers are obtained under idealized assumptions such as the use of the Monte Carlo truth in the event reconstruction, and these values represent the best cases achievable under these detector conditions; nevertheless, they show that a large-scale LArTPC is a promising Compton camera for cosmic MeV gamma-ray observations.

	\section*{Acknowledgment}
	This study was supported by Toray Science and Technology Grant No. 20-6104 (Toray Science Foundation), and by JSPS KAKENHI Grant Numbers 18H05861, 19H01906, 19H05185, 22H00128, 22K18277(HO), and 24KJ0905 (SA).

	\bibliography{nima}

	\end{document}